\documentclass[runningheads]{llncs}
\usepackage[T1]{fontenc}
\usepackage{graphicx}
\usepackage{amsmath} 
\usepackage{booktabs}

\usepackage{color,soul}
\usepackage[hidelinks]{hyperref}

\usepackage{multirow}	
\usepackage{amsfonts}						%
\usepackage{duckuments}						%

\usepackage[table]{xcolor}
\usepackage{listings}
\usepackage{pifont}
\usepackage{pdflscape}
\usepackage{lscape}
\usepackage{array}
\usepackage{comment}
\usepackage{rotating}
\usepackage{enumitem}
\usepackage{dirtree}
\usepackage{wrapfig}
\usepackage{subcaption}

\usepackage{float}

\usepackage{caption}

\usepackage{graphicx}

\usepackage{listings,amsmath}
\usepackage{pifont}
\usepackage{pdflscape}
\usepackage{multirow}
\usepackage{lscape}
\usepackage{graphicx}
\usepackage{array}
\usepackage{comment}
\usepackage{rotating,booktabs}

\definecolor{codegreen}{rgb}{0,0.6,0}
\definecolor{codegray}{rgb}{0.5,0.5,0.5}
\definecolor{codepurple}{rgb}{0.58,0,0.82}
\definecolor{backcolour}{rgb}{0.95,0.95,0.92}

\lstdefinestyle{mystyle}{
    backgroundcolor=\color{backcolour},   
    commentstyle=\color{codegreen},
    keywordstyle=\color{magenta},
    numberstyle=\tiny\color{codegray},
    stringstyle=\color{codepurple},
    basicstyle=\ttfamily\footnotesize,
    breakatwhitespace=false,         
    breaklines=true,                 
    captionpos=b,                    
    keepspaces=true,                 
    numbers=left,                    
    numbersep=5pt,                  
    showspaces=false,                
    showstringspaces=false,
    showtabs=false,                  
    tabsize=2
}

\usepackage{fnbreak}

\usepackage[hidelinks]{hyperref}

\usepackage[misc]{ifsym}

\begin{document}

\title{Bridging RDF Knowledge Graphs with Graph Neural Networks for Semantically-Rich Recommender Systems}

\titlerunning{Bridging RDF KGs with GNNs for Semantically-Rich Recommender Systems}

 \author{Michael Färber~\Letter\inst{1}\textsuperscript{\orcid{0000-0001-5458-8645}}%
 \and David Lamprecht\inst{2}\textsuperscript{\orcid{0000-0002-9098-5389}}
 \and Yuni Susanti\inst{3}\textsuperscript{\orcid{0009-0001-1314-0286}}}

\hyphenation{Sem-Open-Alex}
\hyphenation{data-sets}

\newcommand{\orcid}[1]{\href{https://orcid.org/#1}{\includegraphics[width=10pt]{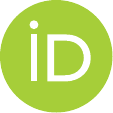}}}

\authorrunning{M. Färber et al.} %

\institute{ScaDS.AI \& TU Dresden, Dresden, Germany\\
\email{michael.faerber@tu-dresden.de}
\and
metaphacts GmbH, Walldorf, Germany\\\email{dl@metaphacts.com}
\and
Fujitsu Ltd., Japan\\\email{susanti.yuni@fujitsu.com}
}

\maketitle  

\begin{abstract}
Graph Neural Networks (GNNs) have substantially advanced the field of recommender systems. However, despite the creation of more than a thousand knowledge graphs (KGs) under the W3C standard RDF, their rich semantic information has not yet been fully leveraged in GNN-based recommender systems. To address this gap, we propose a comprehensive integration of RDF KGs with GNNs that utilizes both the topological information from RDF \textit{object} properties and the content information from RDF \textit{datatype} properties. Our main focus is an in-depth evaluation of various GNNs, analyzing how different semantic feature initializations and types of graph structure heterogeneity influence their performance in recommendation tasks. Through experiments across multiple recommendation scenarios involving multi-million-node RDF graphs, we demonstrate that harnessing the semantic richness of RDF KGs significantly improves recommender systems and lays the groundwork for GNN-based recommender systems for the Linked Open Data cloud. The code and data are available on our GitHub repository.\footnote{\url{https://github.com/davidlamprecht/rdf-gnn-recommendation}} 

\end{abstract}

\section{Introduction}
\label{sec:introduction}
\vspace{-0.1cm}
Recommender systems have become essential tools for alleviating information overload and enhancing user experience in a wide range of applications~\cite{guo2020survey,wu2022graph}. Graph Neural Network (GNN)-based recommendation methods have gained significant attention due to their capability to effectively process structured data and leverage high-order information~\cite{gao2023survey}. Research has shown that GNN-based models consistently outperform traditional approaches across various benchmark datasets~\cite{wu2022graph}, highlighting their growing importance in the field.

\begin{figure}[tb] %
\centering 
\includegraphics[width=0.9\textwidth]{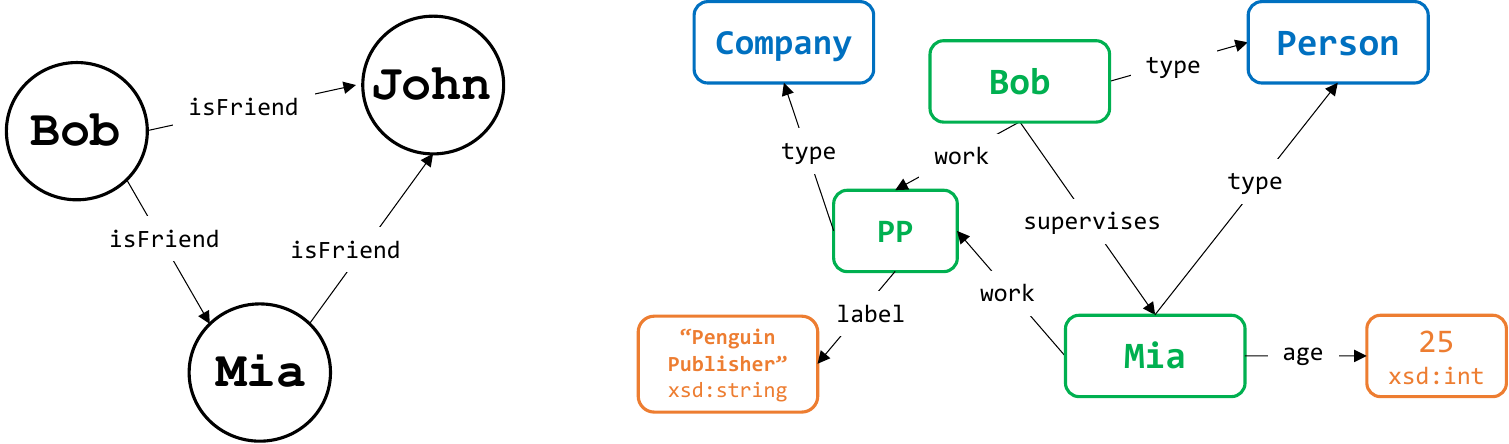}
\caption{Illustration of homogeneous (left) vs. heterogeneous (right) graphs (RDF).}
\label{fig:hethom}
\end{figure}

However, despite the growing prominence of GNN-based methods in recommender systems, their application has largely focused on \textit{homogeneous} graphs and user-item recommendation scenarios. 
In a homogeneous graph (see Figure~\ref{fig:hethom}), all nodes are of the same type, and all edges represent the same kind of relationship. For instance, in a \textit{social network} graph, nodes represent individuals, and edges represent their connections. Other examples include \textit{product} graphs and \textit{citation networks}~\cite{wu2022graph,zhou2020graph,hu2020open}. 
In contrast, a \textit{heterogeneous} graph, such as knowledge graphs (KGs), includes diverse types of nodes and edges. KGs are typically represented as sets of triples using standard formats like the Resource Description Framework (RDF)~\cite{w3Primer}.
RDF is a W3C standard that serves as the basis for more than thousand knowledge graphs in the Linked Open Data cloud \cite{PolleresKFTM20}, including Wikidata, YAGO, and DBpedia \cite{FarberBMR18}, as well as domain-specific ones like SemOpenAlex~\cite{semopenalex}, STRING \cite{szklarczyk2023string} and ChEMBL \cite{zdrazil2024chembl} in the biomedical domain. These graphs often contain millions or billions of entities and relationships and thus far exceed the order of magnitude that is usually used in evaluation benchmarks for GNNs. Furthermore, these graphs contains not only multiple entity and relation types, but also RDF \textit{datatype} properties (e.g., \texttt{xsd:string} in Figure~\ref{fig:hethom}). In addition, the rich semantic relationships among nodes in KGs-based recommender systems enhance node representation~\cite{Wang_2018} and improve the interpretability of recommendation results~\cite{YANG2020106194}. 

When dealing with heterogeneous graphs such as RDF KGs, existing approaches often fall short of fully leveraging their semantic potential due to their complex structure with multi-type entities and relations~\cite{wu2022graph}.
Thus, current GNN methods struggle to fully process and utilize the comprehensive \textit{semantics} of RDF, which includes the graph topological information and content-based information derived from various types of literals. 
To address these challenges, we present a GNN-RDF-based recommendation system leveraging both the RDF's topological and content-based information as semantically-rich features,
further
laying the groundwork for Linked Open Data-based recommendation with many readily-available RDF KGs. Our approach bridges the gap between the Graph Machine Learning community (e.g., ICLR, LOG, ICML) and the Semantic Web community (e.g., ISWC, ESWC, CIKM), eliminating the need for semantic technologies such as SPARQL--the query language for RDF KGs.
Our approach utilizes \texttt{AutoRDF2GML}~\cite{autordf2024}, a framework to convert RDF data 
into heterogeneous graph data suitable for GNNs.  
The main contribution of this work is the evaluation of how semantic features and graph structure heterogeneity in RDF KGs affect the performance of GNN architectures across multiple recommendation scenarios. Our evaluation on two large RDF KGs, comprising up to 22 million RDF triples, demonstrates that semantic feature-based initialization techniques and the incorporation of heterogeneous graph structures can significantly boost performance. 
Our work also provides a solid basis for selecting GNN-based recommender system for 
RDF data. Overall, we make the following contributions:
\begin{enumerate}[leftmargin=0.5cm]
   \item We present a GNN-based recommendation pipeline aligned with RDF knowledge graphs. Our approach leverages the comprehensive semantics of RDF data,
   including both the RDF object properties (topological information) and RDF datatype properties (content-based information).
   \item We evaluate the impact of various semantic feature initializations and graph structure heterogeneity on the performance of prominent GNN architectures. 
   Our evaluation on two large RDF knowledge graphs across multiple recommendation scenarios shows that leveraging the semantic depth of RDF data significantly improves the performance of GNN-based recommender systems.

\end{enumerate}

\vspace{-0.7cm}
\section{Framework} %
\label{sec:rdfgnn}
\vspace{-0.3cm}
Our approach is composed of two main modules: (1) \textbf{RDF to Heterogeneous Graphs Transformation} (\S\ref{sec:autordf}) and (2) \textbf{GNN-based Recommendation Systems} (\S\ref{sec:gnndriven}), explained in the following. 
\vspace{-0.3cm}

\vspace{-0.2cm}
\subsection{RDF to Heterogeneous Graphs Transformation }
\label{sec:autordf}
\vspace{-0.1cm}
To automatically transform large RDF data into a heterogeneous graph dataset, \texttt{AutoRDF2GML}~\cite{autordf2024}, a comprehensive framework to convert RDF data into heterogeneous graph datasets suitable for graph-based machine learning methods, has been recently published. \texttt{AutoRDF2GML} automatically extracts numeric features from the RDF knowledge graph by utilizing both content-based information and topology-based relationships. A key advantage of \texttt{AutoRDF2GML} is its ability to automatically select and transform these semantically-rich content-based features, making it accessible to users without expertise in RDF and SPARQL.

\begin{figure}[tb] %
\centering 
\includegraphics[width=0.90\textwidth]{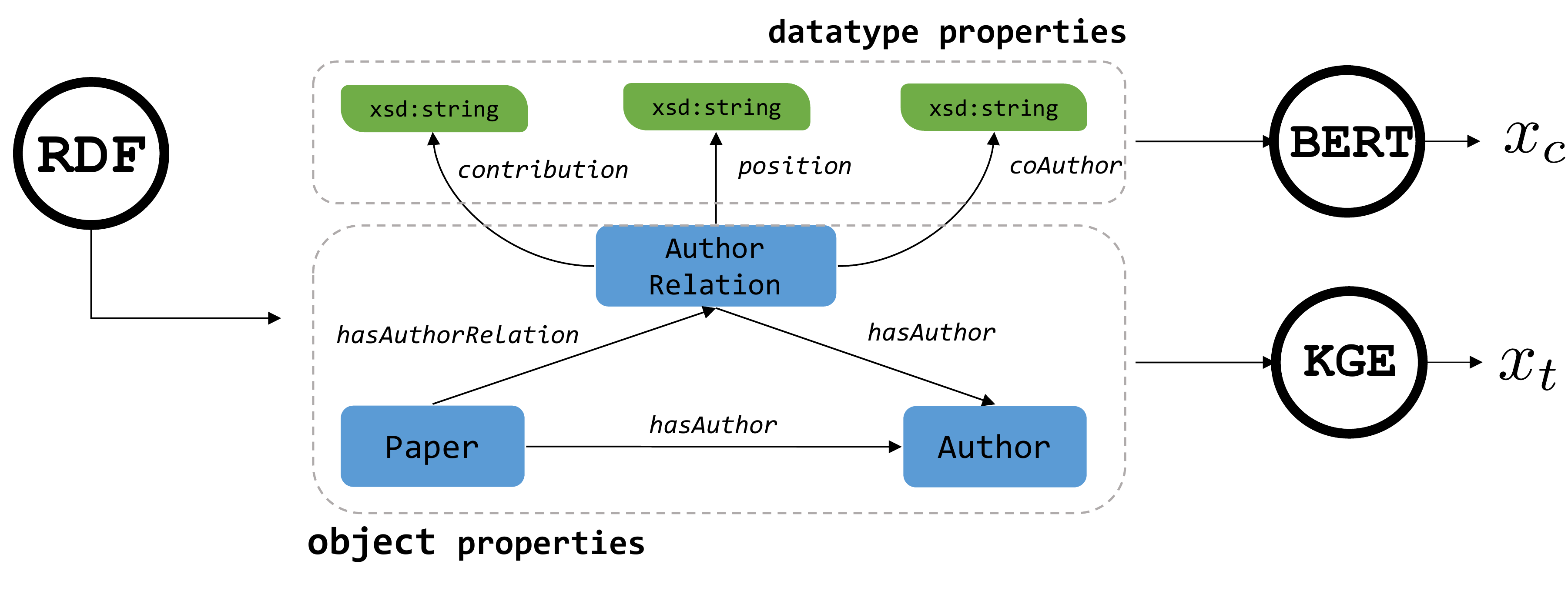}
\caption{Illustration of automatic features extraction with \texttt{AutoRDF2GML}.} 
\label{fig:autordf}
\end{figure}

Figure~\ref{fig:autordf} illustrates the features extraction process with \texttt{AutoRDF2GML}. 
Given an RDF knowledge graph $\mathcal{KG}$, we transform it into a heterogeneous graph dataset $\mathcal{HG}$, with 
$\mathcal{KG}\mapsto \mathcal{HG}(x_t,x_c)$. This dataset $\mathcal{HG}$ includes topology-based features $x_t$ and content-based features $x_c$, as follows:
\begin{itemize}[leftmargin=0.4cm]
    \item \textbf{Content-based features $x_c$} : We utilize \textbf{RDF datatype properties} to capture the text semantic information of RDF knowledge graphs as content-based features. 
    For example, \texttt{``contribution''} and \texttt{``coAuthor''} in Figure~\ref{fig:autordf} are RDF datatype properties. Depending on the type of literal, \texttt{AutoRDF2GML} applies specific transformation rules. For instance, \textit{string} literals are transformed into numerical features using BERT~\cite{devlin-etal-2019-bert}-based embedding models.
    \item \textbf{Topology-based features $x_t$} : We leverage the RDF knowledge graph's topological information through its \textbf{RDF object properties} as the topology-based features. For example, \texttt{``hasAuthor''} in Figure~\ref{fig:autordf} is an RDF object property.  Techniques for obtaining these representations include Knowledge Graph Embedding (KGE) techniques such as TransE \cite{bordes2013translating}, DistMult \cite{yang2014embedding}, ComplEx \cite{trouillon2016complex}, and RotatE \cite{sun2019rotate}, which effectively encode RDF entities into feature vectors. \texttt{AutoRDF2GML} automates this process using these techniques.
\end{itemize}

\vspace{-0.5cm}
\subsection{GNN-based Recommendation Systems with RDF Features}
\label{sec:gnndriven}
\begin{figure}[tb] 
\centering 
\includegraphics[width=1\textwidth]{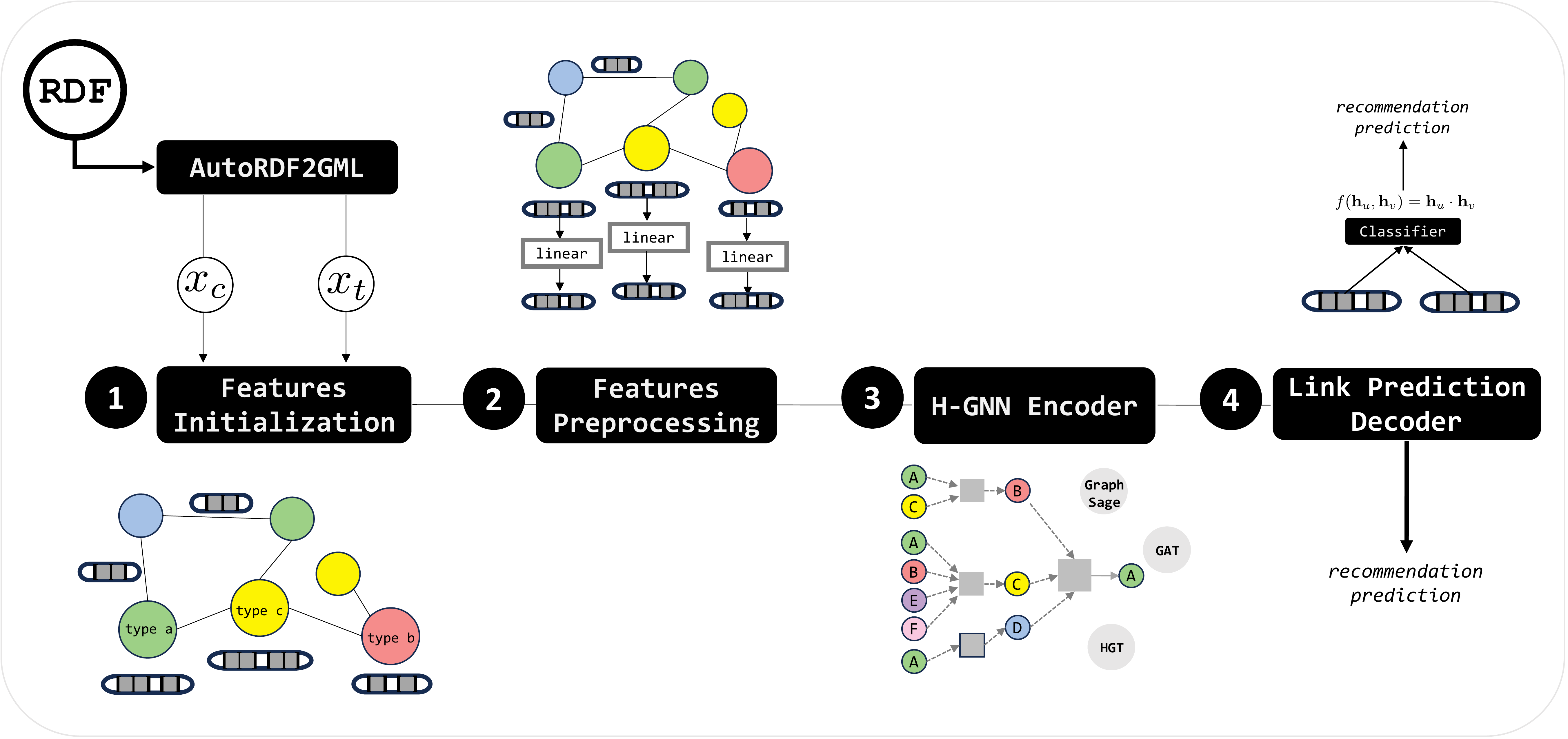}
\caption{GNN-based recommendation system with RDF-based features.} 
\label{fig:GNN-recommendation-pipline}
\end{figure}

\vspace{-0.2cm}
To demonstrate the benefits of integrating semantic features from RDF knowledge graphs into GNN-based recommendation system, we thoroughly evaluate the heterogeneous graph dataset generated in the previous step by integrating it into GNN-based recommender system, illustrated in Figure~\ref{fig:GNN-recommendation-pipline}. Our approach involves an in-depth exploration of various semantic feature initialization methods (\S\ref{sec:semantic-feature-initialization2}) and types of graph heterogeneity, on the performance of prominent GNN architectures
across multiple recommendation scenarios (\S\ref{sec:scenario}). Our primary goal is to analyse how GNNs can exploit semantic node features, further identify the most effective methods for combining these features having different semantic characteristics to enhance GNN-based recommendation systems.
As illustrated in Figure~\ref{fig:GNN-recommendation-pipline}, the GNN-based recommendation pipeline follows these steps:
\begin{enumerate}[leftmargin=0.5cm]
    \item \textbf{Feature Initialization:} Formally, feature initialization step assigns initial feature vectors $x_v^0$ to each node \( v \) in the heterogeneous graph \( \mathcal{HG} = (\mathcal{V}, \mathcal{E}) \), where \( \mathcal{V} \) is the set of nodes and \( \mathcal{E} \) is the set of edges. These initial feature vectors $x_v^0$ are automatically computed e.g. with \texttt{AutoRDF2GML}, correspond to $x_c$ and $x_t$, as explained in \S\ref{sec:autordf}. 
    In this process, our pipeline addresses both graph structure heterogeneity and feature heterogeneity; including the variations in vector dimensionality and underlying semantics, detailed in \S\ref{sec:semantic-feature-initialization2}.
    \item \textbf{Feature Preprocessing:} 
    Due to the heterogeneity of the features extracted from the underlying RDF data, different node types may have features in different spaces. 
    We apply linear transformation $\mathbf{x}_v = \mathbf{W} x_v + \mathbf{b}$ to project the features of varying node types into a unified feature space in a same dimension, following existing works \cite{zhang2019heterogeneous,wang2019heterogeneous,hu2020heterogeneous}.
    \item \textbf{H-GNN Encoder:} We implement several heterogeneous GNN architectures to obtain the contextual node representations \( \mathbf{h}_v \) of the graph. In a GNN, the contextual representation \( \mathbf{h}_v \) of a node \( v \) is computed from its features \( \mathbf{x}_v \) through the GNN function $\mathbf{h}_v = \text{GNN}(x_v)$
where \(\text{GNN}(\cdot)\) denotes the function process the node features and its neighborhood to produce the contextual representation. We detailed our choice of GNN architectures in \S\ref{sec:models}.

    \item \textbf{Link Prediction Decoder:} 
    Finally, a classifier is used for link prediction,
    producing the final recommendation.
    Given a pair of feature vectors $h_u$ and $h_v$ from the encoder, the classifier predicts the probability of existence of an edge by computing a similarity score $\hat{y}_{u \sim v}$ 
    using \textit{dot product} classifier: $\hat{y}_{u \sim v} = f(\mathbf{h}_u, \mathbf{h}_v) = \mathbf{h}_u \cdot \mathbf{h}_v$. We employ dot product due to its simplicity and strong performance, particularly in recommendation scenarios with a large number of potential links~\cite{sagesda}. 
\end{enumerate}
\vspace{-0.5cm}
\subsubsection{Semantic Feature Initialization.}
\label{sec:semantic-feature-initialization2}
\vspace{-0.1cm}
Nodes can be initialized with various feature vectors. In the following, we outline different node feature initializations with different levels of semantic richness. We consider both content-based and topology-based features, as well as their various combinations:
\begin{enumerate}[leftmargin=0.5cm]
    \item \textbf{One-hot-encoding (\texttt{one-hot}).} As a baseline approach, we employ one-hot encoding for feature initialization, a method commonly used for non-attributed graphs \cite{cui2022positional,lv2021we}. Unlike the other forms of feature initialization, one-hot vectors do not encapsulate any semantic information. 
    \item \textbf{Content-based: Natural language description (NLD, \texttt{cb\textsubscript{nld}}).} This initialization method uses text embeddings, such as 
    BERT~\cite{devlin-etal-2019-bert} embeddings, derived from natural language descriptions of entities (e.g., labels, abstracts). 
    \item \textbf{Content-based: Literals (\texttt{cb\textsubscript{Literal}}).~}
    This initialization leverages all available RDF literals for the feature vectors. 
    Beyond the textual descriptions of entities, these RDF literals can be categorical, numeric, boolean, or of another data type. Thus, in contrast to the \textit{NLD} initialization, the initialization vectors are extended by context-based information of all data types. 
    \item \textbf{Topology-based (\texttt{tb}).} 
    This initialization method uses encoded information about the topology of the entities in the input. 
    Such information can be derived using knowledge graph embedding model (e.g., TransE \cite{bordes2013translating}).
    \item \textbf{NLD if available, otherwise topology-based (\texttt{comb\textsubscript{nld|tb}}).} 
    This method initializes with \textit{NLD} vectors when available; otherwise, it uses topology-based embeddings.
    This approach is commonly used in related works (\S\ref{sec:related-work}) when content-based node features are not provided for all node types \cite{yu2020scalable,he2023spectral,yang2023simple,chi2022residual}.
\end{enumerate}

Additionally, we include the combination of content-based and topology-based features to enhance the semantic richness of the node features, \textbf{6. Concatenation (\texttt{comb\textsubscript{Concat}})}, \textbf{7. Addition (\texttt{comb\textsubscript{Addition}})}, \textbf{8. Weighted Addition (\texttt{comb\textsubscript{WAddition}})}, \textbf{9. Average (\texttt{comb\textsubscript{Average}})}, and \textbf{10. Neural Combinator (\texttt{comb\textsubscript{nc}})} utilizes a feedforward neural network $z = f(W \cdot [a, b] + c)$.

\vspace{-0.5cm}
\section{Evaluation}
\label{sec:evaluation}

\vspace{-0.3cm}
\subsubsection{Choice of GNN Architectures.} We employ the following three widely-used models: \textbf{(1) GraphSAGE }\cite{hamilton2017inductive} 
samples and aggregates features from node's local neighborhood, integrating topological and feature information using three aggregators: mean, LSTM, and pooling~\cite{kipf2016semi,zhou2020graph,fey2019fast}, \textbf{(2) Graph Attention Network (GAT)}~\cite{velivckovic2017graph} uses masked self-attention to weigh relevant neighbors, bypassing costly matrix operations and prior graph structure knowledge, while enabling message passing with multidimensional edge features for better node updates~\cite{fey2019fast}, and \textbf{(3) Heterogeneous Graph Transformer (HGT)} \cite{hu2020heterogeneous} is designed for complex heterogeneous graphs, using type-specific parameters and meta-relation-based attention to handle diverse node and edge types while learning implicit metapaths.
\label{sec:models}
While HGT is specifically designed for heterogeneous graphs \cite{hu2020heterogeneous}, GraphSAGE and GAT were originally designed for homogeneous graphs. Thus, we applied GraphSAGE and GAT models to heterogeneous graphs by implementing the message and update functions for each edge type \cite{fey2019fast}. We use default parameters from GNN literature~\cite{lv2021we,hu2020heterogeneous,wang2019neural} for training the GNN models. We provide the detailed implementation settings including hyperparameters settings and other technical details for the model training in our Github.

\subsubsection{Dataset.}
\label{sec:data}
\vspace{-0.2cm}

We evaluate on real-world recommendation tasks using the heterogeneous graph datasets \texttt{SOA-SW}~\cite{soaswbenchmark} and \texttt{LPWC} \cite{lpwcbenchmark}, generated via \texttt{AutoRDF2GML} \cite{autordf2024} from the RDF knowledge graphs \textit{SemOpenAlex} \cite{semopenalex} and \textit{LinkedPapersWithCode} \cite{farber2023linked}. 
\texttt{SOA-SW} consists of 21.9 million RDF triples and models scientific works and their associated entities from the Semantic Web community. \texttt{LPWC} consists of 7.9 million RDF triples and provides comprehensive information about machine learning papers, including the tasks addressed, the datasets utilized and the methods implemented. Compared to other benchmarks that do not utilize RDF data (see Table~\ref{tab:comparisonHeteroGraphBenchmarks}), these datasets are unique in incorporating the semantic features from both topological and content-based information.

\begin{table}[tbp]
    \centering
    \caption{Evaluation of GNN models with varying feature initializations and graph heterogeneity for \textit{\textbf{paper}} recommendation scenario on \texttt{SOA-SW}.}
    \label{tab:eval-soa-sw-author-paper}
    \resizebox{1.0\textwidth}{!}{%
    \begin{tabular}{|c|l|cccc|cccc|cccc|}
    \hline
    & Feature & \multicolumn{4}{c|}{GraphSAGE} & \multicolumn{4}{c|}{GAT} & \multicolumn{4}{c|}{HGT}  \\ 
    & Initialization & F1  & Pre & Re & AUC & F1 & Pre & Re & AUC & F1 & Pre & Re & AUC \\ \hline
    \hline
    \multirow{10}{*}{\rotatebox[origin=c]{90}{Full Heterogenous Graph}} & \texttt{one-hot} & 0.806 & 0.899 & 0.731 & 0.937 & 0.875 & 0.925 & 0.830 & 0.962 & 0.890 & 0.880 & 0.901 & 0.949 \\
    & \texttt{cb\textsubscript{nld}} & 0.874 & 0.932 &0.823~ & 0.967 & 0.877 & 0.924 & 0.834 & 0.961 & 0.886 & 0.901 & 0.872 & 0.957 \\ 
    & \texttt{cb\textsubscript{Literal}} & 0.914 & 0.927 & 0.901 & 0.972 & 0.889 & 0.919 & 0.861 & 0.964 & 0.887 & 0.882 & 0.892 & 0.945 \\
    & \texttt{tb} & 0.926 & 0.956 & 0.899 & 0.983 & 0.910 & 0.942 & 0.880 & 0.975 & 0.915 & 0.935 & 0.896 & 0.976 \\ 
    & \texttt{comb\textsubscript{nld|tb}} & 0.933 & 0.959 & 0.908 & 0.985 & 0.920 & 0.929 & 0.910 & 0.973 & 0.906 & 0.943 & 0.872 & 0.976 \\ 
    & \texttt{comb\textsubscript{Concat}} & 0.931 & 0.951 & 0.911 & 0.982 & 0.918 & 0.948 & 0.890 & 0.979 & 0.925 & \textbf{0.949} & 0.902 & \textbf{0.982} \\ 
    & \texttt{comb\textsubscript{Addition}} & 0.921 & 0.951 & 0.892 & 0.982 & 0.922 & \textbf{0.956} & 0.889 & 0.982 & 0.882 & 0.939 & 0.832 & 0.970 \\ 
    & \texttt{comb\textsubscript{WAddition}} & \textbf{0.940} & 0.950 & \textbf{0.929} & 0.984 & \textbf{0.923} & 0.954 & 0.894 & \textbf{0.983} & 0.885 & 0.934 & 0.841 & 0.968 \\ 
    & \texttt{comb\textsubscript{Average}} & 0.926 & \textbf{0.963} & 0.893 & \textbf{0.987} & 0.898 & 0.932 & 0.866 & 0.971 & \textbf{0.934} & 0.937 & \textbf{0.931} & 0.977 \\ 
    & \texttt{comb\textsubscript{nc}} & 0.896 & 0.941 & 0.855 & 0.973 & 0.889 & 0.867 & \textbf{0.912} & 0.941 & 0.889 & 0.913 & 0.865 & 0.961 \\ \hline \hline
    
    \multirow{10}{*}{\rotatebox[origin=c]{90}{Bipartite Graph}} & \texttt{one-hot} & 0.810 & 0.893 & 0.740 & 0.855 & 0.823 & 0.894 & 0.763 & 0.863 & 0.824 & 0.896 & 0.763 & 0.850 \\ 
    & \texttt{cb\textsubscript{nld}} & 0.855 & 0.911 & 0.805 & 0.942 & 0.830 & 0.871 & 0.793 & 0.892 & 0.854 & 0.836 & \textbf{0.873} & 0.924 \\ 
    & \texttt{cb\textsubscript{Literal}} & 0.882 & 0.910 & 0.856 & 0.955 & 0.846 & 0.852 & 0.841 & 0.903 & 0.847 & 0.846 & 0.848 & 0.914 \\ 
    & \texttt{tb} & \textbf{0.936} & \textbf{0.969} & 0.905 & \textbf{0.987} & \textbf{0.895} & \textbf{0.914} & 0.877 & \textbf{0.952} & \textbf{0.892} & \textbf{0.940} & 0.850 & \textbf{0.967} \\ 
    & \texttt{comb\textsubscript{nld|tb}} & 0.905 & 0.904 & 0.905 & 0.965 & 0.872 & 0.877 & 0.866 & 0.928 & 0.828 & 0.898 & 0.768 & 0.915 \\ 
    & \texttt{comb\textsubscript{Concat}} & 0.922 & 0.936 & \textbf{0.908} & 0.977 & 0.891 & 0.890 & \textbf{0.893} & 0.941 & 0.872 & 0.902 & 0.844 & 0.945 \\ 
    & \texttt{comb\textsubscript{Addition}} & 0.904 & 0.960 & 0.854 & 0.978 & 0.855 & 0.904 & 0.810 & 0.942 & 0.884 & 0.937 & 0.837 & 0.963 \\ 
    & \texttt{comb\textsubscript{WAddition}} & 0.910 & 0.956 & 0.869 & 0.977 & 0.873 & 0.902 & 0.845 & 0.939 & 0.876 & 0.904 & 0.850 & 0.949 \\ 
    & \texttt{comb\textsubscript{Average}} & 0.906 & 0.949 & 0.867 & 0.973 & 0.876 & 0.888 & 0.865 & 0.940 & 0.866 & 0.875 & 0.857 & 0.933 \\ 
    & \texttt{comb\textsubscript{nc}} & 0.849 & 0.906 & 0.800 & 0.939 & 0.846 & 0.890 & 0.807 & 0.924 & 0.818 & 0.918 & 0.738 & 0.915 \\ \hline
    \end{tabular}
    } %
    \vspace{-3mm}
\end{table}

\vspace{-0.3cm}
\subsubsection{Recommendation Scenarios.}
\label{sec:scenario}
In GNN-based recommendations, bipartite or homogeneous graphs are often used, which include only the node types relevant to the recommendation.
To evaluate the effect of graph heterogeneity, we compare the performance of the recommender system on such graphs with using full heterogeneous graph, on the following recommendation scenarios:
\begin{enumerate}[leftmargin=0.5cm]
\item \textbf{Paper Recommendation}:~
In this scenario, we perform link prediction on \textit{author}-\textit{work} edge using the \texttt{SOA-SW} dataset, experimenting with both full heterogeneous graph setting (6 node types, 7 edge types) and bipartite graph setting containing \textit{author} and \textit{work} nodes and their edges (\textit{author}-\textit{work}).

\item \textbf{Collaboration Recommendation}:~ 
In this scenario, we perform link prediction on the \textit{author}-\textit{author} edge type using the \texttt{SOA-SW} dataset. Since this scenario involves only \textit{author} nodes, we experiment with both homogeneous and heterogeneous graph settings: the heterogeneous setting uses the entire graph, while the homogeneous setting focuses on \textit{author} nodes and \textit{author}-\textit{author} edges forming the \textit{co-author} network. 

\item \textbf{Task Recommendation}:~ 
In this scenario, we perform link prediction on the \textit{dataset}-\textit{task} edge type using the \texttt{LPWC} dataset, on both full heterogeneous graph (4 node types, 6 edge types) and bipartite graph setting focusing on \textit{dataset} and \textit{task} nodes and their connecting edges.
\end{enumerate}

\vspace{-0.5cm}
\section{Results and Discussions}
\label{sec:resultdis}
\vspace{-0.2cm}
The evaluation results for \textit{paper}, \textit{collaboration}, and \textit{task} recommendation scenarios are summarized in Table~\ref{tab:eval-soa-sw-author-paper}, \ref{tab:eval-soa-sw-author-author}, and \ref{tab:eval-lpwc-dataset-task}, respectively. Following existing works~\cite{wu2022graph,wang2019knowledgerec}, we use the following evaluation metrics: ROC-AUC score (AUC), F1-score (F1), Precision (Pre), and Recall (Re). The best performance on \textit{paper} recommendation scenario was achieved by GraphSAGE using the full heterogeneous graph setting: with feature initialization method \texttt{comb\textsubscript{WAddition}}, it reached an F1-score of 0.940, and with method \texttt{comb\textsubscript{Average}}, it obtained the highest ROC-AUC value of 0.987.
Similarly, for \textit{collaboration} recommendation scenario, the best performances were achieved by GraphSAGE in the full heterogeneous graph setting: with \texttt{comb\textsubscript{WAddition}}, it achieved the highest F1-score of  \textbf{0.804}, and with \texttt{comb\textsubscript{Average}}, the highest ROC-AUC value of \textbf{0.969}. 
In \textit{task} recommendation scenario, similar to the other two scenarios, the best performance was achieved by GraphSAGE in the full heterogeneous graph setting (highest F1-score of \textbf{0.923} with \texttt{comb\textsubscript{Average}}; highest ROC-AUC value of \textbf{0.975} with \texttt{comb\textsubscript{WAddition}}). 
In the following, we discuss our results from several perspectives.
\begin{table}[tbp]
    \centering
    \caption{Evaluation of GNN models with varying feature initializations and graph heterogeneity for \textit{\textbf{collaboration}} recommendation scenario on \texttt{SOA-SW}.}
    \label{tab:eval-soa-sw-author-author}
    \resizebox{1.0\textwidth}{!}{%
    \begin{tabular}{|c|l|cccc|cccc|cccc|}
    \hline
    & Feature & \multicolumn{4}{c|}{GraphSAGE} & \multicolumn{4}{c|}{GAT} & \multicolumn{4}{c|}{HGT}  \\ 
    & Initialization & F1  & Pre & Re & AUC & F1 & Pre & Re & AUC & F1 & Pre & Re & AUC \\ \hline
    \hline
    \multirow{10}{*}{\rotatebox[origin=c]{90}{Full Heterogenous Graph}} & \texttt{one-hot} & 0.787 & 0.661 & 0.973 & 0.940 & 0.790 & 0.668 & 0.966 & 0.938 & 0.744 & 0.633 & 0.903 & 0.871 \\
    & \texttt{cb\textsubscript{nld}} & 0.793 & 0.665 & 0.984 & 0.957 & 0.797 & 0.671 & 0.983 & 0.956 & 0.773 & 0.654 & 0.946 & 0.894 \\ 
    & \texttt{cb\textsubscript{Literal}} & 0.795 & 0.666 & 0.987 & 0.952 & 0.744 & 0.601 & 0.977 & 0.861 & 0.751 & 0.642 & 0.905 & 0.884 \\
    & \texttt{tb} & 0.791 & 0.663 & 0.979 & 0.921 & 0.800 & 0.676 & 0.981 & 0.947 & 0.780 & 0.654 & 0.966 & 0.911 \\ 
    & \texttt{comb\textsubscript{nld|tb}} & 0.803 & \textbf{0.677} & 0.989 & 0.960 & \textbf{0.803} & 0.676 & 0.988 & 0.952 & 0.743 & 0.613 & 0.941 & 0.869 \\ 
    & \texttt{comb\textsubscript{Concat}} & 0.801 & 0.676 & 0.982 & 0.950 & 0.798 & 0.675 & 0.976 & 0.941 & \textbf{0.794} & \textbf{0.665} & \textbf{0.987} & \textbf{0.959} \\ 
    & \texttt{comb\textsubscript{Addition}} & 0.795 & 0.669 & 0.980 & 0.946 & 0.802 & 0.676 & 0.985 & 0.958 & 0.767 & 0.645 & 0.946 & 0.890 \\ 
    & \texttt{comb\textsubscript{WAddition}} & \textbf{0.804} & \textbf{0.677} & 0.989 & 0.961 & \textbf{0.803} & 0.675 & \textbf{0.990} & \textbf{0.964} & 0.760 & 0.632 & 0.956 & 0.886 \\ 
    & \texttt{comb\textsubscript{Average}} & 0.803 & 0.675 & \textbf{0.991} & \textbf{0.969} & 0.801 & 0.675 & 0.984 & 0.949 & 0.765 & 0.641 & 0.950 & 0.890 \\ 
    & \texttt{comb\textsubscript{nc}} & 0.796 & 0.666 & 0.989 & 0.942 & 0.802 & \textbf{0.677} & 0.985 & 0.949 & 0.736 & 0.611 & 0.926 & 0.870 \\ \hline \hline
    
    \multirow{9}{*}{\rotatebox[origin=c]{90}{Homogeneous Graph}} & \texttt{one-hot} & 0.739 & 0.644 & 0.869 & 0.844 & 0.755 & 0.663 & 0.876 & 0.868 & - & - & - & - \\ 
    & \texttt{cb\textsubscript{nld}} & 0.741 & 0.639 & 0.881 & 0.848 & 0.756 & 0.663 & 0.879 & 0.885 & - & - & - & - \\
    & \texttt{cb\textsubscript{Literal}} & 0.771 & \textbf{0.672} & 0.904 & 0.906 & 0.757 & 0.673 & 0.865 & 0.877 & - & - & - & - \\
    & \texttt{tb} & 0.774 & 0.663 & \textbf{0.928} & 0.909 & 0.751 & 0.649 & \textbf{0.892} & \textbf{0.888} & - & - & - & - \\ 
    & \texttt{comb\textsubscript{Concat}} & \textbf{0.776} & 0.668 & 0.927 & \textbf{0.913} & 0.757 & 0.666 & 0.877 & 0.885 & - & - & - & - \\
    & \texttt{comb\textsubscript{Addition}} & 0.765 & 0.659 & 0.914 & 0.902 & 0.757 & 0.667 & 0.875 & 0.879 & - & - & - & - \\ 
    & \texttt{comb\textsubscript{WAddition}} & 0.762 & 0.656 & 0.910 & 0.898 & \textbf{0.764} & \textbf{0.679} & 0.874 & 0.882 & - & - & - & - \\ 
    & \texttt{comb\textsubscript{Average}} & 0.762 & 0.657 & 0.908 & 0.898 & 0.745 & 0.644 & 0.882 & 0.837 & - & - & - & - \\ 
    & \texttt{comb\textsubscript{nc}} & 0.741 & 0.641 & 0.879 & 0.834 & 0.736 & 0.634 & 0.879 & 0.825 & - & - & - & - \\ \hline
    \end{tabular}
    } %
    \vspace{-3mm}
\end{table}
\vspace{-0.5cm}
\subsubsection{Analyis of Semantic Feature Initialization Methods.}
Across all our experiments (Tables 5-7), feature initialization techniques using semantic node features consistently outperformed those with one-hot vectors. Additionally, our analysis showed that heterogeneous graph structures are superior to both bipartite and homogeneous graph configurations. Our findings highlight the effectiveness of using \texttt{comb\textsubscript{WAddition}} and \texttt{comb\textsubscript{Average}} as the most successful combination methods. These approaches integrate semantic information from both content and topological structure of the underlying RDF data, giving the best results.

\vspace{-0.5cm}
\subsubsection{Performance across Different GNN Architectures.}
In all three experiments, GraphSAGE, when initialized with either \texttt{comb\textsubscript{WAddition}} or \texttt{comb\textsubscript{Average}} and applied to the full heterogeneous graph setting, consistently outperforms other models across all evaluation metrics. These results suggest that GraphSAGE is particularly effective at leveraging semantically rich node features, as also shown by \cite{hamilton2017inductive,kipf2016semi}.
Additionally, this result aligns with \cite{lv2021we}, which showed that GNN architectures designed for homogeneous graphs (e.g., GraphSAGE) can surpass more complex heterogeneous graph neural networks (HGNNs) when provided with appropriate input. This proves the importance of semantically rich node features in the GNN-based recommendation tasks.
Conversely, in scenarios where one-hot vectors without semantic information are used as node features, GAT and HGT consistently outperform GraphSAGE in the full heterogeneous graph setting. This is shown in Tables~\ref{tab:eval-soa-sw-author-paper}--\ref{tab:eval-lpwc-dataset-task}, where GraphSAGE achieves F1 scores of 0.806, 0.787, and 0.739, while GAT/HGT achieves scores of 0.89, 0.79, and 0.778, respectively. This underscores their effectiveness in leveraging the graph's structure, even when semantically rich node features are not available~\cite{hu2020heterogeneous,velivckovic2017graph,lv2021we}.
\begin{table}[tbp]
    \centering
    \caption{Evaluation of GNN models with varying feature initializations and graph heterogeneity for \textit{task} recommendation on \texttt{LPWC}.}
    \label{tab:eval-lpwc-dataset-task}
    \resizebox{1.0\textwidth}{!}{%
    \begin{tabular}{|c|l|cccc|cccc|cccc|}
    \hline
    & Feature & \multicolumn{4}{c|}{GraphSAGE} & \multicolumn{4}{c|}{GAT} & \multicolumn{4}{c|}{HGT}  \\ 
    & Initialization & F1  & Pre & Re & AUC & F1 & Pre & Re & AUC & F1 & Pre & Re & AUC \\ \hline
    \hline
    \multirow{9}{*}{\rotatebox[origin=c]{90}{Heterogeneous Graph}} & \texttt{one-hot} & 0.739 & 0.791 & 0.694 & 0.834 & 0.748 & 0.766 & 0.732 & 0.794 & 0.778 & 0.801 & 0.756 & 0.859 \\
    & \texttt{cb\textsubscript{nld}} & 0.837 & 0.873 & 0.803 & 0.928 & 0.802 & 0.742 & 0.872 & 0.847 & 0.783 & 0.721 & 0.857 & 0.862 \\ 
    & \texttt{cb\textsubscript{Literal}} & 0.784 & 0.697 & 0.895 & 0.853 & 0.800 & 0.834 & 0.769 & 0.879 & 0.820 & 0.778 & 0.868 & 0.894 \\
    & \texttt{tb} & 0.903 & 0.913 & 0.893 & 0.965 & 0.868 & \textbf{0.916} & 0.826 & 0.936 & 0.877 & 0.837 & 0.922 & 0.936 \\ 
    & \texttt{comb\textsubscript{Concat}} & 0.907 & 0.926 & 0.889 & 0.970 & 0.873 & 0.911 & 0.839 & 0.936 & 0.826 & 0.779 & 0.878 & 0.898 \\ 
    & \texttt{comb\textsubscript{Addition}} & 0.912 & 0.921 & 0.903 & 0.967 & 0.875 & 0.890 & 0.860 & 0.936 & \textbf{0.885} & \textbf{0.845} & \textbf{0.930} & \textbf{0.943} \\ 
    & \texttt{comb\textsubscript{WAddition}} & 0.918 & \textbf{0.933} & 0.903 & \textbf{0.975} & 0.872 & 0.896 & 0.849 & 0.938 & 0.875 & 0.841 & 0.912 & 0.936 \\ 
    & \texttt{comb\textsubscript{Average}} & \textbf{0.923} & 0.920 & 0.926 & 0.971 & \textbf{0.882} & 0.882 & 0.883 & \textbf{0.942} & 0.829 & 0.767 & 0.903 & 0.896 \\ 
    & \texttt{comb\textsubscript{nc}} & 0.879 & 0.832 & \textbf{0.932} & 0.943 & 0.825 & 0.766 & \textbf{0.894} & 0.886 & 0.783 & 0.811 & 0.757 & 0.875 \\ 
    \hline \hline    
    \multirow{10}{*}{\rotatebox[origin=c]{90}{Bipartite Graph}} & \texttt{one-hot} & 0.782 & 0.762 & 0.803 & 0.851 & 0.737 & 0.601 & \textbf{0.951} & 0.745 & 0.697 & 0.541 & 0.980 & 0.743 \\ 
    & \texttt{cb\textsubscript{nld}} & 0.836 & 0.800 & 0.876 & 0.903 & 0.806 & 0.771 & 0.845 & 0.852 & 0.795 & 0.718 & 0.891 & 0.868 \\ 
    & \texttt{cb\textsubscript{Literal}} & 0.756 & 0.739 & 0.773 & 0.823 & 0.815 & 0.779 & 0.855 & 0.859 & 0.798 & 0.739 & 0.868 & 0.869 \\
    & \texttt{tb} & \textbf{0.915} & 0.930 & 0.901 & 0.971 & \textbf{0.830} & \textbf{0.809} & 0.852 & \textbf{0.898} & 0.819 & \textbf{0.866} & 0.776 & \textbf{0.912} \\ 
    & \texttt{comb\textsubscript{Concat}} & 0.904 & \textbf{0.936} & 0.873 & \textbf{0.973} & 0.799 & 0.717 & 0.903 & 0.869 & \textbf{0.845} & 0.803 & 0.892 & 0.909 \\ 
    & \texttt{comb\textsubscript{Addition}} & 0.889 & 0.902 & 0.876 & 0.955 & 0.803 & 0.750 & 0.864 & 0.878 & 0.721 & 0.574 & \textbf{0.971} & 0.772 \\ 
    & \texttt{comb\textsubscript{WAddition}} & 0.882 & 0.836 & \textbf{0.933} & 0.940 & 0.791 & 0.717 & 0.882 & 0.865 & 0.794 & 0.726 & 0.875 & 0.854 \\ 
    & \texttt{comb\textsubscript{Average}} & 0.862 & 0.885 & 0.841 & 0.940 & 0.793 & 0.726 & 0.872 & 0.869 & 0.732 & 0.599 & 0.939 & 0.769 \\ 
    & \texttt{comb\textsubscript{nc}} & 0.813 & 0.746 & 0.895 & 0.875 & 0.740 & 0.617 & 0.923 & 0.725 & 0.740 & 0.603 & 0.958 & 0.767 \\ \hline
    \end{tabular}
    } %
    \vspace{-1mm}
\end{table}
\vspace{-0.9cm}
\subsubsection{Full Heterogeneous vs. Bipartite Graph Settings.} 
Our findings emphasize that in bipartite graph settings, feature initialization with TransE knowledge graph embeddings--a topology-based feature--consistently delivers the best results, as shown by the performance metrics in the lower half of Tables~\ref{tab:eval-soa-sw-author-paper} and \ref{tab:eval-lpwc-dataset-task}. 
Specifically, in the \textit{paper} recommendation scenario on \texttt{SOA-SW} (see Table~\ref{tab:eval-soa-sw-author-paper}), the best-performing full heterogeneous graph seting using \texttt{comb\textsubscript{WAddition}} feature initialization achieved an F1-score of 0.940. In comparison, the bipartite graph setting initialized with TransE embeddings attained an F1-score of 0.936, showing a minimal difference of just 0.4\%. Similarly, for \textit{task} recommendation scenario on \texttt{LPWC} (see Table~\ref{tab:eval-lpwc-dataset-task}), the best full heterogeneous graph seting with \texttt{comb\textsubscript{Average}} achieved an F1-score of 0.923, while the bipartite graph setting using TransE embeddings resulted in an F1-score of 0.915, a difference of 0.8\% score.

\vspace{-0.3cm}
\section{Related Work}
\label{sec:related-work}

\vspace{-0.3cm}
\subsubsection{GNNs for Heterogeneous Graphs and the Benchmarks.}
Heterogeneous graphs are used to model different types of objects and relationships 
\cite{AnsarizadehTTR23,wu2022graph}, offering a more accurate representation of real-world phenomena compared to homogeneous graphs.
This structured modeling improves application performance, with GNNs outperforming other methods in recommender systems \cite{gao2023survey}. 
In Table~\ref{tab:comparisonHeteroGraphBenchmarks}, we summarized benchmark datasets for heterogeneous graph tasks. 
\begin{wraptable}[13]{r}{5cm}\vspace*{-2\baselineskip}%
\tiny
 \centering 
\caption{Heterogeneous Graphs}
\label{tab:comparisonHeteroGraphBenchmarks}
\vspace{-2mm}
\begin{tabular}{lccc}
\toprule
\textbf{Benchmarks} & \texttt{NLD} & \texttt{Literals$\setminus$NLD} & \texttt{Topology} \\
\midrule
AMiner \cite{dong2017metapath2vec} & \ding{55} & \ding{55} & \ding{55} \\
MovieLens \cite{movielens} & \ding{55} & \ding{55} & \ding{55} \\
LastFM \cite{fu2020magnn} & \ding{55} & \ding{55} & \ding{55} \\
HGB\_Freebase \cite{lv2021we} & \ding{55} & \ding{55} & \ding{55} \\
HGB\_LastFM \cite{lv2021we} & \ding{55} & \ding{55} & \ding{55} \\
Douban \cite{zhang2019inductive} & \ding{55} & \ding{55} & \ding{55} \\
Flixster \cite{zhang2019inductive} & \ding{55} & \ding{55} & \ding{55} \\
Yahoo-Music \cite{zhang2019inductive} & \ding{55} & \ding{55} & \ding{55} \\
OGB\_MAG \cite{hu2020open} & \ding{51} & \ding{55} & \ding{55} \\
DBLP \cite{fu2020magnn} & \ding{51} & \ding{55} & \ding{55} \\
IMDB \cite{fu2020magnn} & \ding{51} & \ding{55} & \ding{55} \\
HGB\_DBLP \cite{lv2021we} & \ding{51} & \ding{55} & \ding{55} \\
HGB\_ACM \cite{lv2021we} & \ding{51} & \ding{55} & \ding{55} \\
HGB\_PubMed \cite{lv2021we} & \ding{51} & \ding{55} & \ding{55} \\
HGB\_Amazon \cite{lv2021we} & \ding{51} & \ding{51} & \ding{55} \\
HGB\_IMDB \cite{lv2021we} & \ding{51} & \ding{51} & \ding{55} \\
\midrule   
SOA-SW \cite{soaswbenchmark} 
& \ding{51} & \ding{51} & \ding{51} \\
LPWC \cite{lpwcbenchmark}
& \ding{51} & \ding{51} & \ding{51} \\
\bottomrule
\end{tabular}
\end{wraptable}
Table~\ref{tab:comparisonHeteroGraphBenchmarks} shows that, apart from \texttt{SOA-SW} and \texttt{LPWC}, most benchmark datasets do not use RDF data. They often lack node features for all node types, or when present, these features mainly stem from inherent NLD properties. 
Consequently, performance gains on these benchmarks may result from improved topology-based node features rather than true advancements in GNN models, which reflect optimized feature engineering~\cite{li2023long}. As of this writing, \texttt{SOA-SW} and \texttt{LPWC} are the only large graph datasets that provide both content-based (\texttt{NLD} and \texttt{Literals$\setminus$NLD}) and topology-based node features for all node types, ideal for our evaluation.

\vspace{-0.5cm}
\subsubsection{GNNs for Recommender Systems.} GNNs are widely used in recommender systems because these systems often involve graph-structured data~\cite{gcnweb10.1145/3219819.3219890,berg2017graphconvolutionalmatrixcompletion}, and GNNs excel at graph representation learning~\cite{wu2022graph}. Studies show that GNN-based models outperform traditional~\cite{guo2020survey,wu2022graph} and prior GNN-based 
methods~\cite{neuralgra10.1145/3331184.3331267,gcnweb10.1145/3219819.3219890}.
The advantage of GNNs especially lies in their ability to explore multi-hop relationships, which has been shown to significantly benefit recommender systems~\cite{neuralgra10.1145/3331184.3331267}. 
Recent advancements in Large Language Models (LLMs) have introduced their application in recommender systems, with approaches like~\cite{llmrec10.1145/3523227.3546767} framing recommendation problems as prompt-based natural language tasks. However, employing prompt-based methods on graph-structured data presents challenges, as it requires transforming complex graph content into textual prompts. In addition, research shows that LLMs and GNNs often complement each other effectively~\cite{abs-2406-01032}. For that reason, we focus on GNNs-based methods in the current work.

\vspace{-0.3cm}
\section{Conclusion}
\label{sec:conclusion}
\vspace{-0.3cm}
In this paper, we integrate RDF knowledge graphs with Graph Neural Networks (GNNs), demonstrating the advantages of using the rich 
semantics of RDF to improve GNN-based recommendation systems. We conducted an extensive evaluation across multiple recommendation scenarios, analyzing the impact of various semantic node feature initializations and graph structure heterogeneity. Our results demonstrate that these components significantly enhance the performance of GNN-based recommender systems.
Additionally, 
by considering RDF knowledge graphs with their global knowledge, our approach
not only uses the node types directly relevant to recommendations, but also auxiliary nodes that provide valuable additional contextual information. 
This approach effectively facilitates the creation of semantically enriched features from both content and topological information, leading to improved performance of GNN-based recommender systems.
In future work, we will consider the combination of several RDF knowledge graphs from the Linked Open Data cloud for GNN-based recommendation.

\bibliographystyle{splncs04}
\bibliography{reference,references}
\end{document}